\documentstyle[12pt]{article}
\begin{document}
\baselineskip=21pt

\hfill {  SU-HEP-4240-551 }

\hfill {  September 1993 }
\vskip 1.0truecm
\centerline{\bf VIOLATION OF THE $\displaystyle{\Delta I = {1\over 2}} $
 RULE IN $\displaystyle{D\rightarrow
\pi\pi}$ DECAYS}
\vskip 1cm
\centerline{ Francesco Sannino$^{(a,b)}$ }
\vskip 1cm
\centerline{\it (a) Department of Physics, Syracuse University,}
\centerline{\it Syracuse,  New York, 13244-1130 }
\vskip 1cm
\centerline{\it (b) INFN, Sezione di Napoli, Mostra d'Oltremare Pad.16,}
\centerline{\it I-80125, Napoli, Italy}
\vskip 3cm
\centerline{\bf ABSTRACT}
\vskip 1cm
A strong violation of the $\displaystyle{
\Delta I = {1\over 2}} $ rule has experimentally
 been found
 in the  $\displaystyle{D \rightarrow \pi\pi}$ decays [1]. In this letter
we will show that the
order of magnitude of this
violation can be understood in terms of the pure quantum chromo dynamics
corrections to the weak interactions.

\newpage

$\bullet$ {\bf Introduction}
\vskip 2cm

The approximate $\displaystyle{\Delta I={1\over2}}$ rule was firstly
established in the
kaon decays and
 until now never experimentally violated.

The strong $\displaystyle{\Delta I={3\over2}}$ suppression is still now not
completely
 understood and so it remains one of the more interesting and open problems
in elementary particle physics.
It was suggested [2] that one possible solution to this puzzle could be
the coupling of the strong interactions to the weak currents. This idea for
the kaon decays cannot quantitatively work as QCD under $1~GeV$ is not
perturbative and relatively big soft gluon corrections should be expected.
Nevertheless it gives us a non trivial physical picture in order to
understand the
 phenomenological rule.
The suppression in this scheme is only due to the underlying color dynamics
of the interacting gluons and quarks.
Now if we assume that the gluon exchanges (hard and soft) really affect the
weak currents, we expect a {\it dynamical} correction, in other words, the
 corrections should depend on the energy scale involved in the
decays.
It is really the different energy scale between the $D$ and $K$ decays which
must play a fundamental role in the understanding of the
 $\displaystyle{\Delta I={1\over2}}$ violation in $D$ decays.
It is also possible to give a {\it naive } physical picture in which as the
 c mass is greater than the s mass we expect that the dynamical gluon
corrections
don't strongly affect the weak operators for the $D$ decays.
In this letter we haven't used the standard factorization approach
which is based on the
valence-quark assumption and vacuum-insertion approximation. In that scheme the
matrix elements of two quark bilinear operators are saturated by the vacuum
intermediate states in all possible ways.
We believe that even if the factorization scheme has some
suggestive features it actually breaks the
fundamental symmetries of the whole effective weak operators, leading to
some wrong results [11,14].
On the contrary, we assume that the total symmetries of the QCD
regularized bilinear weak operators are preserved by the hadronization and
this in turns allows us to
easily relate the $D\rightarrow \pi\pi$ weak amplitudes to each other.

\vfill\eject
\newpage
$\bullet $ {\bf Experimental starting point}

\vskip 2cm

The usual isospin decomposition for the $D \rightarrow \pi\pi$ decays can be
written as in the kaon decays [1,4,5]

$$A^{+0}=\sqrt{{3\over2}} A_2 e^{i\delta_2}$$

$$A^{+-}={1\over {\sqrt{3}}} A_2 e^{i\delta_2}+
\sqrt{{2\over3}} A_0 e^{i\delta_0}\eqno(1.1)$$

$$A^{00}=\sqrt{{2\over3}} A_2 e^{i\delta_2}-
{1\over {\sqrt{3}}} A_0 e^{i\delta_0}$$
\vskip 1cm
where $A^{+0}$, $A^{+-}$ and $A^{00}$ are the amplitudes for $D^{+}\rightarrow
\pi^{+}\pi^{0}$,  $D^{0}\rightarrow \pi^{+}\pi^{-}$ and  $D^{0}\rightarrow
\pi^{0}\pi^{0}$ respectively.

In this standard notation the
$\delta_2$ and $\delta_0$ phases are related to
the strong final state interactions of $\pi\pi$ elastic scattering.
This decomposition, as well known, is based on the Watson's theorem [5,6] which
uses
 the unitarity condition and the elastic structure of the strong
interaction for the {\it in} and {\it out} states.
If we relax the condition for the elastic structure of the final state
interactions, the decomposition (1.1) can still be made, but
$\delta_2$ and $\delta_0$ should just be interpreted as new parameters.
Some author [11] uses complex phases
to parametrize a possible strong $\pi\pi$ inelastic final state interaction.
Here we will assume elastic $\pi\pi$ final state interactions.
The experimental analysis based on (1.1) leads to the following
very interesting results [1]:

$$\left|{{{A_2} \over{A_0}}}\right|
=0.72\pm 0.13 \pm 0.11\eqno(1.2)$$

and

$$\cos (\delta_2 - \delta_0)=0.14 \pm 0.13 \pm 0.09\eqno(1.3) $$

\vskip .7cm
As the $A_2$ amplitude is comparable to the
$A_0$ amplitude, the $\displaystyle{\Delta I={1\over2}}$ rule,
established in the
$K\rightarrow \pi\pi$ decays, is badly broken here.
Now we are going to understand this different behaviour in terms of
the underlying
quantum chromo dynamics.

\vfill\eject
\newpage

$\bullet$ {\bf Dynamical behaviour of the $\Delta I = {1\over 2}$ rule}

\vskip 2cm

The bare weak non-leptonic hamiltonian which contributes to
$D\rightarrow \pi\pi$
is [3,10]:

$$H_W={{G_F}\over {\sqrt{2}}}\sin(\vartheta_c)\cos(\vartheta_c)
\left[(\overline{d}u)(\overline{c}d)+
(\overline{u}d)(\overline{d}c)\right]\eqno(2.1)$$
\vskip .7cm
where $(\overline{q}_2 q_1)$ denotes an uncolored left $(V-A)$ current and
$\vartheta_c$
 is the Cabibbo angle.
Taking into account hard gluon exchange we have the effective non leptonic
weak hamiltonian [2,3,10]:

$$H_{W}^{eff.}={{G_F}\over {2\sqrt{2}}}\sin(\vartheta_c)\cos(\vartheta_c)
\left[ a_{+}I_{+}+a_{-}I_{-}\right]+h.c.\eqno(2.2)$$
where
$$I_{\pm}=(\overline{u}d)(\overline{d}c)\pm (\overline{d}d)(\overline{u}c)
\eqno(2.3)$$
\vskip .7cm
$a_{+}$ and $a_{-}$ are coefficients which summarize the underlying
dynamics of the hard gluon-quark interactions. Their explicit expression
 is

$$a_{\pm}=\left({{\alpha_{S}(m_c)}\over{\alpha_{S}(M_W)}}\right)^
{\gamma_{\pm}}\eqno(2.4)$$
where
$$\gamma_{-}=-2\gamma_{+}={4\over b}~~~~~~~~~~
b=11-{2\over 3}N_f~~~~~~~~~~~~~N_f=4\eqno(2.5)$$
\vskip .7cm
and $\gamma_{\pm}$ are called the anomalous dimensions.
The energy scale dependence of the strong coupling constant is

$$\alpha_{S}(Q^2)={{4\pi}\over {b \ln \left({{Q^2}\over{\Lambda^2}}\right)}}
\eqno(2.6)$$
\vskip .7cm
The standard notation has been used here. We use $\alpha_{S}(M_W)\cong 0.113$
 [17], while for $\alpha_{S}(m_c)$, due to the big uncertainty on $\Lambda$,
we use two different values: $\alpha_{S}(m_c)\cong 0.2, 0.3$.
 It is easy show that

$$a_{+}\cong 0.87~~~~~~~~~~a_{-}\cong 1.31$$
$$\eqno(2.7)$$
$$a_{+}\cong 0.79~~~~~~~~~~a_{-}\cong 1.60$$

\vskip .5cm
\noindent
where the the first and second line are respectively due to
$\alpha_{S}(m_c)\cong 0.2$ and $\alpha_{S}(m_c)\cong 0.3$.
The new effective operators $I_{\pm}$ have well
defined behaviours under isospin transformations.
Indeed, we will see that $I_{-}$ is a pure
$\displaystyle{\Delta I={1\over 2}}$
operator while $I_{+}$ is a mixture of $\displaystyle{\Delta I={3\over 2}}$ and
$\displaystyle{\Delta I={1\over 2}}$ [5].
Looking at the $a_{\pm}$ coefficients we observe a relatively small dynamical
suppression of the $\displaystyle{\Delta I= {3\over2}}$ operator so that we
 expect a failure of the $\displaystyle{\Delta I= {1\over 2}}$ rule in
 the $D$ decays.
The crucial point is that at the $m_c$ scale this perturbative picture
should work
 better than at the kaon scale as we will show.
We will not consider the penguin diagrams, which are fundamental
in the kaon decays, because due to
SU(3) flavour symmetry they shouldn't play any fundamental role in the $D$
decays [5,10].

\vfill\eject
\newpage
$\bullet $ {\bf Theoretical Results and Conclusions}
\vskip 2cm

To make quantitative estimates we will use the isospin properties of the
$I_{\pm}$ operators and assume that the subsequent hadronization actually
preserves the whole non-dynamical information carried by the effective weak
operators.
Let us expand the $I_{\pm}$ operators in terms of operators with definite
isospin:

$$I_{+}={2\over \sqrt{3}}~T^{3\over2}_{1\over 2} +
{1\over \sqrt{6}}~\hat{T}^{1\over2}_{1\over 2} -
{1\over \sqrt{2}}~T^{1\over2}_{1\over 2}\eqno(3.1)$$

$$I_{-}=
\sqrt{{3\over 2}}~\hat{T}^{1\over2}_{1\over 2}+
{1\over \sqrt{2}}~T^{1\over2}_{1\over 2}\eqno(3.2)$$
\vskip .7cm
As claimed $I_{-}$ is a pure $\displaystyle{\Delta I={1\over 2}}$ operator
while $I_{+}$ is
a combination of $\displaystyle{\Delta I={3\over 2}}$ and $
\displaystyle{\Delta I={1\over 2}}$.
The two different $\displaystyle{\Delta I={1\over 2}}$ operators
$T^{1\over2}_{1\over 2}$ and
$\hat{T}^{1\over2}_{1\over 2}$ are due respectively to the following products
of two SU(2) isospin representations
$\displaystyle{0\otimes{1\over2}}$ and $\displaystyle{1\otimes{1\over2}}$.
Now using the Bose symmetrized $\pi\pi$ final states and the Wigner-Eckart
theorem
we will be able to express the ratio
$\displaystyle{\left|{{{A_2} \over{A_0}}}\right|}$
as a function of the dynamical
coefficients  $a_{\pm}$ and of the reduced matrix elements.
The three independent reduced matrix elements are defined as:

$$M_2:=<2\| T^{3\over 2} \|1/2>$$
$$M_0:=<0\| T^{1\over 2} \|1/2>\eqno(3.3)$$
$$\hat{M}_0:=<0\| \hat{T}^{1\over 2} \|1/2>$$
\vskip .7cm
It's finally possible to obtain the following model independent
expression

$$\left|{{{A_2} \over{A_0}}}\right|=2 a_{+}\sqrt{2}
\left|{{M_2}\over {\hat{M}_0}}\right|~
\left| {(a_{+}+3a_{-})+\sqrt{3}{{M_0}\over {\hat{M}_0}}(a_{-}-a_{+})}\right|^
{-1}\eqno(3.4)$$
\vskip .7cm

To obtain quantitative results we need a scheme to deduce the three reduced
matrix elements $M_2$, $M_0$ and $\hat{M}_0$. Unfortunately, we still
do not have a reliable model for
the non leptonic matrix elements and that is why after almost 30 years
 the $\displaystyle{\Delta I = {3\over 2}} $ suppression in
$K\rightarrow 2\pi$ is still unsatisfactorily explained.
On the contrary, we can see
 that, due to different energy scales, the simplest ansatz on those
reduced matrix elements is able to explain the new experimental data
  for $D\rightarrow
2\pi$.
Our hypothesis for the reduced matrix elements is that the
non perturbative strong interactions which are involved in the subsequent
hadronization do not break the total effective operators symmetries.
So we simply assume:

$$M_{2}=M_{0}=\hat{M}_{0}\eqno(3.5)$$
\vskip .7cm
We can than rewrite (3.4) in terms only of the $a_{\pm}$

$$\left|{{{A_2} \over{A_0}}}\right|={{2 a_{+}\sqrt{2}}\over{
\left| {(a_{+}+3a_{-})+\sqrt{3}(a_{-}-a_{+})}\right|}}\eqno(3.6)$$
\vskip .7cm
Using the numerical expressions for $a_{\pm}$ we have for
$\alpha_{S}(m_c)\cong 0.3$  and $\alpha_{S}(m_c)\cong 0.2$ respectively:

$$\left|{{{A_2} \over{A_0}}}\right|\cong 0.32,~~0.44\eqno(3.8)$$
\vskip .7cm
We can see comparing the theoretical results with the experimental one, which
is $\displaystyle{
\left|{{{A_2} \over{A_0}}}\right|=0.72\pm 0.13\pm0.11}$, that this simple
model gives the right order of magnitude and that the theoretical value is
about
two standard
deviations
 from the experimental one. We also note that experimentally the
$\displaystyle{\Delta I={3\over 2}}$ is more enhanced than the theoretical
prediction and that
in turn should reflect on the reduced matrix elements.
On the contrary if we use the same idea for the $K$
decays, we recall that a too large theoretical
$\displaystyle{\Delta I={3\over2}}$
 is expected [2]. One possible explanation of this interesting puzzle
could be a
little
inelastic $\pi\pi$ scattering in the $D$ decays.
We can also have soft-gluon effects, and these are expected, roughly
speaking, at the $D$ scale to be of order $~~\displaystyle
{{{\Lambda_{QCD}}\over{m_c}}\cong 14\%}$ with $\displaystyle{\Lambda_{QCD}
\cong 200~MeV}$ while we need a correction about $40\%$ to obtain the
central experimental value.
Another little contribution could came from non-expected penguin operators
which
should enhance the $\displaystyle{\Delta I={1\over2}}$ amplitude.

However, to really disentangle this problem we need new more precise
 measurements of
the ratio $\displaystyle{\left|{{{A_2} \over{A_0}}}\right|}$.
We can conclude observing that
our theoretical result indicates that the observed violation of the
$\displaystyle{\Delta I={1\over 2}}$ rule for $D\rightarrow \pi\pi$
is effectively due to the perturbative underlying
chromo dynamics,
which can be identified with the dynamical coefficients $a_{\pm}$.

\vfill\eject
\newpage
\centerline{\bf Acknowledgments}
\vskip 1cm
It's a great pleasure to thank Prof. Giancarlo Moneti which suggested to me
 this very interesting problem and Prof. Joseph Schechter for helpful
discussions, and for careful reading of the manuscript.
\vfill\eject
\newpage

\vskip 1cm

\centerline{\bf References}

\vskip 1cm

\noindent
[1]  CLEO collaboration, CLNS 93/1225 CLEO 93-07, to be published.

\noindent
[2]  M.K. Gaillard and B.W. Lee, Phys. Rev. Lett. $\underline{33}$, 108(1974)

\noindent
{}~~~~~G. Altarelli and L. Maiani, Phys. Lett. B 52, 351(1974)

\noindent
{}~~~~~N. Cabibbo and L. Maiani, Phys. Lett. B 73, 418(1978)

\noindent
[3]  F.J. Gilman and M.B. Wise, Phys. Rev. D $\underline{20}$, 2392(1979)

\noindent
[4]  E.D. Commins, Weak Interactions, McGraw Hill, New York, (1973)

\noindent
[5] L. Okun, Quark e Leptoni, Editori Riuniti, (1986)

\noindent
[6] J. Donoghue, E. Golowich and  B.R. Holstein, Dynamics of the Standard
Model, Cambridge University Press

\noindent
[7]  V. Barger and S. Pakvasa, Phys. Rev. Lett. $\underline{43}$, 812(1979)

\noindent
[8]  G. Eilam and J.P.Leveille, Phys. Rev. Lett. $\underline{44}$, 1648(1980)

\noindent
[9]  J. Finjord, Nucl. Phys. B 181, (1981)74-90

\noindent
[10] B. Yu Blok and M.A. Shifman, Sov. J. Nucl. Phys. {\bf 45}(1), 135(1987)

\noindent
[11] L.L. Chau and H.Y. Cheng, Phys. Rev. D $\underline{36}$, 137(1987)

\noindent
[12] Heavy Quark Physics, Conference Proceedings AIP 196, ITHACA, NY 1989,
Editors: Persis S., Drell and L. Rubini.

\noindent
[13]  M. Gronau and D. London, Phys. Rev. Lett. $\underline{65}$, 3381(1990)

\noindent
[14] J. Liu, Modern Phys. Lett. A 29, (1991)2693-2696.

\noindent
[15] A. Czarnecki, A.N. Kamal and Q. Xu, Z. Phys. C {\bf 54}, 411(1992)

\noindent
[16] L.L. Chau and H.Y. Cheng, Phys. Lett. B 280, (1992) 281-286

\noindent
[17] Particle Data Group, K. Hikasa et al., Phys. Rev. D
$\underline{45}$,(1992)

\noindent
[18] A.J. Buras et al. Nucl. Phys. B 370, (1992)69-104

\noindent
[19] F. Buccella et al. Phys. Lett. B 302, (1993) 319

\noindent
[20] K. Terasaki and S. Oneda, Phys. Rev. D. $\underline{47}$,199(1993)
\vfill \eject

\end {document}